\documentstyle[11pt]{article}
\hsize \textwidth      
\advance \hsize by -\marginparwidth
\evensidemargin \oddsidemargin  

\advance\hoffset by 5mm 

\newtheorem{thm}{Theorem}

\newtheorem{lemma}{Lemma}

\newcommand{\Rm}{I\!\!R}
\newcommand{\Le}{\hbox{Le}}
\newcommand{\vz}{{\bf z}}
\newcommand{\vx}{{\bf x}}
\newcommand{\ve}{{\bf e}}
\newcommand{\eps}{\varepsilon}
\newcommand{\pdr}[2]{\frac{\partial{#1}}{\partial{#2}}}
\begin{document}
\title{An upper bound for the bulk burning rate for systems}

%Enhancement of the traveling front speeds in passive Reaction-diffusion 
%in the presence of advection.    

\author{{Alexander Kiselev  and  Leonid Ryzhik}\\
Department of Mathematics \\ University of Chicago\\
Chicago IL 60637}

\date{}
\maketitle
\begin{abstract}
  We consider a system of reaction-diffusion equations with passive
  advection term and Lewis number $\Le$ not equal to one.  Such
  systems are used to describe chemical reactions in a flow in a situation
  where temperature and material diffusivities are not equal.  It is
  expected that the fluid advection will distort the reaction front,
  increasing the area of reaction and thus speeding up the reaction
  process.  While a variety of estimates on the influence of the flow
  on reaction  are available for a single reaction-diffusion equation
  (corresponding to the case of Lewis number equal to one), the case
  of the system is largely open.  We prove a general upper bound on
  the reaction rate in such systems in terms of the reaction rate
  for a single reaction-diffusion equation, showing that the long time
  average of reaction rate with $\Le \ne 1$ does not exceed the
  $\Le=1$ case. Thus the upper estimates derived for $\Le=1$ apply to
  the systems. Both front-like and compact initial data (hot blob) are
  considered.
\end{abstract}

\section{Introduction}

Systems of reaction-diffusion-advection equations describe numerous
physical processes that occur when reactants not only diffuse but
are also advected by a fluid or wind motion. The simplest situation is
when the effect of temperature and concentration variations on the
fluid motion may be neglected. Then evolution of temperature
$T(t,\vx)$ and concentration $n(t,\vx)$ may be described by a system
of two equations
\begin{eqnarray}
  \label{eq:2.1}
 && T_t+u\cdot\nabla T=\kappa\Delta T+\frac{v_0^2}{\kappa}g(T)n\\
&& n_t+u\cdot\nabla n=\frac{\kappa}{\hbox{Le}}\Delta n-
\frac{v_0^2}{\kappa}g(T)n,~~\vx=(x,y),\nonumber
\end{eqnarray}
where the velocity $u$ is passive and presumed given. 
The nonlinearity $g(T)$ is assumed to be of the KPP-type:
\[
g(0)=0,~~g'(0)\ne 0,~~g(T)\le g'(0)T~~~\hbox{for $T>0$.}
\]
The Lewis number $\Le$, which is the ratio of temperature and material
diffusivities may be different from one. The special case
$\hbox{Le}=1$ was studied in a great detail in the absence of
advection beginning with the pioneering papers of Kolmogorov, Petrovski 
and Piskunov \cite{KPP} and Fisher
\cite{Fisher}. In this case for the initial data satisfying 
$T+n=1$ the system (\ref{eq:2.1}) reduces to a single equation 
for the temperature 
\begin{equation}\label{one}
 T_t+u\cdot\nabla T=\kappa\Delta T+\frac{v_0^2}{\kappa}f(T),
\end{equation}    
where $f(T)=g(T)(1-T).$ Recently the role of advection $u$ in
(\ref{one}) has been a subject of active research (see
\cite{Xin-review} for a review, and
\cite{Vulpiani,ABP,BH,CKOR,Hamel,HPS,KR,CKR} for very recent papers).
In particular, existence of traveling fronts in shear flows \cite{BN}
and pulsating traveling fronts in periodic flows was established
\cite{jxin-2,BH}. It was also shown that asymptotically front-like and
compactly supported data propagate with the speed of the traveling
fronts \cite{BH}. The flow may have drastic effect on the front
propagation, speeding up the reaction. The physical reason for this
phenomenon is believed to be that the advection distorts the front,
helping the hot material to warm up the cold one and increasing the
area available for reaction.  Various estimates for the speed of the
traveling front in the presence of advection when $\Le=1$ were
obtained in \cite{Vulpiani,ABP,CKOR,CKR,KR,HPS}, indicating the
dependence of the burning enhancement on the intensity and geometry of
the flow.

An obvious restriction placed by the model (\ref{one}) is the
assumption ${\rm Le} =1$. There are some interesting situations where
this is clearly not the case, such as a majority of reactions taking
place in viscous liquids, or nuclear combustion in stars, where
temperature diffusivity is much higher than material diffusivity and
one can assume with a good degree of approximation that ${\rm Le} =
\infty.$ It is of interest, therefore, to compare the results already
derived for (\ref{one}) with what one can expect for the system case,
where ${\rm Le} \ne 1.$ The problem turns out to be much harder, in
particular because of the lack of maximum principle for $T.$
Surprisingly little is known about the system (\ref{eq:2.1}) when
$\Le\ne 1$ even in the absence of advection. While it is known that
traveling fronts exist, the set of allowed velocities may differ
significantly from the single equation case, at least for some
(non-KPP) reactions. A. Bonnet \cite{Bonnet} provided examples of
reactions for which (\ref{one}) has a unique traveling front,
(\ref{eq:2.1}) has two disjoint intervals of possible traveling front
velocities.  Furthermore, it is not known that temperature $T$ remains
uniformly bounded in time. The best known estimate \cite{CX} for
$\|T(t)\|_{\infty}$ valid for $u=0$ grows like $\log\log t$ for large $t$.
The main purpose of this paper is to establish an upper bound on the
bulk reaction rate for the full system (\ref{eq:2.1}). It shows that
the rate of reaction for the system is bounded from above by the
traveling front speed for the case $\hbox{Le}=1$. Therefore $\Le\ne 1$
does not provide a speed-up of reaction. This result is established
both for front-like and compactly supported initial data.  We notice
that the papers \cite{CKOR,HPS} contain non-trivial upper bounds on
the speed of front propagation in some cellular flows and in shear flows
respectively. As a corollary of this paper, these bounds extend to the
system case.

For the front problem we consider the system (\ref{eq:2.1}) in a strip
$D={\Rm}_x\times [0,H]_y$. The boundary conditions are periodic in $y$
\begin{equation}\label{eq:bc-per}
T(x,y+H)=T(x,y).
\end{equation}
We assume that initially material on the left is burned, while on the right it
is unburned:
\begin{eqnarray}
  \label{eq:2.2}
T(0,x,y)=T_0(x,y)\to 1,~~n(0,x,y)=n_0(x,y)\to 0~~\hbox{as $x\to -\infty$}\\
T_0(x,y)\to 0,~~n_0(x,y)\to 1~~\hbox{as $x\to +\infty$}.\nonumber
\end{eqnarray}
Initially temperature and concentration 
%satisfy the balance
%\begin{equation}\label{eq:t0+n0=1}
%T_0(x,y)+n_0(x,y)=1,~~~0\le T_0,n_0\le 1
%\end{equation}
%and the function $T_0$ 
are equal to zero and one outside a finite interval:
\begin{equation}\label{compact}
n_0(x,y)=0,\,\,
T_0(x,y)=1~~\hbox{for $x\le -L_0$ and $n_0(x,y)=1,\,\,
T_0(x,y)=0$ for $x\ge L_0$}
\end{equation}
for some $L_0>0$, and
\begin{equation}\label{eq:t0+n0=1}
T_0(x,y)+n_0(x,y) \leq C,~~~ T_0\geq 0,\,\,1\geq n_0\geq 0
\end{equation}
in $[-L_0,L_0].$ 
The flow $u(x,y)\in C^1(\Rm^2)$ is periodic in $x$ and $y$:
\begin{equation}\label{eq:flow-periodic}
u(x+L,y)=u(x,y),~~u(x,y+H)=u(x,y)
\end{equation}
and incompressible:
\[
\nabla \cdot u=0.
\]
We assume in addition that it has mean zero:
\[
\int\limits_0^Ldx\int\limits_0^Hdy u(x,y)=0.
\]

We define the reaction rate as
\[
V(t)=\int\limits_D T_t(x,y)\frac{dxdy}{H}
\]
and its time average as
\begin{equation}\label{eq:time-aver}
\langle V\rangle_t=\frac{1}{t}\int\limits_0^t V(s)ds.
\end{equation}
Long-time propagation speed is described by 
\[ \langle V \rangle_\infty = \limsup_{t \rightarrow \infty} 
\langle V \rangle_t. \]
In a way similar to \cite{CKOR} one may show that, because of the boundary
conditions, we have
\[
V(t)=-\int\limits_D n_t(x,y)\frac{dxdy}{H}=\frac{v_0^2}{\kappa}\int\limits_D
g(T(t,x,y))n(t,x,y)\frac{dxdy}{H}.
\]
It is known \cite{BH} that if $\Le=1$ then there exist pulsating
traveling waves of the form $U(x-ct,x,y)$, periodic in the second two
variables and monotonic in the first. Such solutions exist for $c\ge
c_*$, and solutions with the initial data such as above propagate
asymptotically with the minimal speed $c_*$ \cite{Freidlin,FG}. Our main
result for the front-like data is the following theorem.
\begin{thm}\label{thm1} Let $c_*$ be the minimal speed of the
  pulsating traveling wave for $\Le=1$. Let $T(t,x,y)$, $n(t,x,y)$ be
  solution of (\ref{eq:2.1}) with periodic boundary conditions
  (\ref{eq:bc-per}) and front-like initial data, as in (\ref{eq:2.2}),
  (\ref{eq:t0+n0=1}), (\ref{compact}), with arbitrary Lewis number
  $\Le$. Then we have 
%for $t\ge\frac{\kappa}{v_0^2}$
\[
\langle V\rangle_\infty \le c_*.
\]
%with the constant $C$ depending on the initial data $T_0$.
\end{thm}
\it Remark. \rm In fact we prove that for any $\eps>0,$ 
there exists a finite constant $C_\eps$ such that for any $t,$ 
\[ \langle V \rangle_t \leq c_* +\eps +\frac{C_\eps}{t}. \]

Theorem~\ref{thm1} implies
that difference in material and thermal diffusivities may
not speed up the front propagation relative to the case $\Le=1$.

Our second result applies to the situation when we have initially an
isolated hot spot of material. Let $T_0$ be compactly supported and
$n_0$ satisfy (\ref{eq:t0+n0=1}). We no longer impose periodic
boundary conditions (\ref{eq:bc-per}) but rather consider the problem
in the whole space. Then the reaction rate is defined by
\begin{equation}
  \label{eq:burn-rate-2}
  V(t)=\int\limits_{\Rm^2}T_t(x,y)\frac{dxdy}{H}.
\end{equation}
Its time average is now scaled not as in (\ref{eq:time-aver}) but rather as
\[
\langle\langle V\rangle\rangle_t=\frac{1}{t^2}\int\limits_0^tV(s)ds
\]
and 
\[ \langle\langle V \rangle\rangle_\infty = \limsup_{t \rightarrow \infty}
\langle\langle V\rangle\rangle_t, \]
since the total area of an expanding blob of hot material grows as $t^2$.
We may also define
a traveling front going in direction $\ve$ as a solution of 
\[
T_t+u\cdot\nabla T=\kappa\Delta T+\frac{v_0^2}{\kappa}g(T)(1-T)
\]
of the form $U(\ve\cdot\vx-ct,\vx)$ monotonic in the first variable and
periodic in the second and such that
\[
\lim_{s\to-\infty}U(s,\vx)=1,~~
\lim_{s\to +\infty}U(s,\vx)=0.
\]
Such solutions exist for $c\ge c_*(\theta)$ with $c_*(\theta)$ being the
minimal speed in direction $\theta$.
\begin{thm}\label{thm2} Let $T(t,x,y)$, $n(t,x,y)$ be
  solution of (\ref{eq:2.1}) with the initial $T_0$ that is compactly
  supported and $n_0$ that satisfies (\ref{eq:t0+n0=1}). Then the
long-time average reaction rate satisfies the upper bound
\[
\langle\langle V\rangle\rangle_\infty \le
\frac{1}{2}\int\limits_0^{2\pi}c_{*}^2(\theta)d\theta.
\]
\end{thm}
The first term in the above inequality is equal to the asymptotic bulk
reaction rate of a solution of (\ref{eq:2.1}) with $T_0$ having compact
support and $\Le=1$. Thus Theorem \ref{thm2} is the analog of Theorem
\ref{thm1} for such initial data.

Finally, we note that our arguments also provide some information 
on the important case of ignition-type nonlinearity, where 
the function $g(T)$ in (\ref{eq:2.1}) satisfies $g(T)=0$ 
for $0 \leq T \leq T_0$ for some ``ignition temperature''
$T_0<1.$ Directly from the proofs, it is clear that in this case
we get an upper bound on the reaction rate in terms of the minimal 
traveling wave velocity of the single equation (\ref{one}) with 
any KPP reaction $f(T)$ satisfying $f(T) \geq g(T)(1-T).$ 
\begin{thm}\label{thm3}
% Let $c_*$ be the minimal speed of the
%  pulsating traveling wave for $\Le=1$.
Assume that the reaction function $g(T)$ in (\ref{eq:2.1})
is of ignition type.
 Let $T(t,x,y)$, $n(t,x,y)$ be
  solution of (\ref{eq:2.1}) with periodic boundary conditions
  (\ref{eq:bc-per}) and front-like initial data, as in (\ref{eq:2.2}),
  (\ref{eq:t0+n0=1}), (\ref{compact}), with arbitrary Lewis number
  $\Le$. Let $f(T)$ be any reaction of KPP type
(that is, positive on $(0,1)$ and satisfying 
$f'(0)>0$) such that $f(T) \geq g(T)(1-T).$ Let $c_*$ be the minimal 
speed of the traveling wave in (\ref{one}) with such $f(T).$     
Then we have 
%for $t\ge\frac{\kappa}{v_0^2}$
\[
\langle V\rangle_\infty \le c_*
\]
%with the constant $C$ depending on the initial data $T_0$.
\end{thm}
Our methods also establish existence of a classical solution to
(\ref{eq:2.1}) in a way similar to \cite{CX} extending the result of
that paper to non-zero advection.

We present the proof of Theorem \ref{thm1} in the rest of the paper.
The proof of Theorem \ref{thm3} follows along the same argument.  The main
difficulty in the proof lies in the absence of known uniform
$L^\infty$ bounds on temperature $T$. If such bounds were available
the proof would be greatly simplified. The proof of Theorem \ref{thm2}
is very similar and hence is omitted.

\section{Outline of the proof of Theorem \ref{thm1}}

The proof of Theorem \ref{thm1} will proceed as follows. First we will
show that the front cannot move to the right faster than $c_*$:
\begin{lemma}\label{lemma1} For any $\eps>0$ there exists a constant 
$C_\eps>0$ and a constant $\lambda_\eps>0$  such that 
\[
T(t,x,y)\le C_\eps e^{-\lambda_\eps(x-(c_*+\eps)t)}.
\]
\end{lemma}
However, it is not yet known whether $T(t,x,y)$ is uniformly bounded in
time. Therefore we may not conclude from Lemma \ref{lemma1} that the
conclusion of Theorem \ref{thm1} holds since a lot of reaction may
occur behind the front. To the best of our knowledge the strongest
$L^\infty$-bound on $T$ even for $u=0$ is
\[
T(t,x,y)\le C(1+\hbox{loglog} t)
\]
obtained in \cite{CX}. It suffices for us to establish a weaker upper
bound in the presence of non-zero advection.
\begin{lemma}\label{lemma2} For any $\eps>0$ there exists a
  constant $C_\eps>0$ such that
\[
0\le T(t,x,y)\le C_\eps e^{\eps t}.
\]
\end{lemma}
The proof of Lemma \ref{lemma2} essentially follows the ideas of
\cite{CX} with some modifications. Lemmas \ref{lemma1} and
\ref{lemma2} allow us to prove the analog of Lemma 1 for $n(t,x,y)$:
\begin{lemma}\label{lemma3} 
For any $\eps>0$ there exists a function $\Psi_\eps(\xi) \geq 0$ such that
\[
n(t,x,y)\ge 1-\Psi_\eps{(x-(c_*+\eps)t)}
\]
and
\[
\int_0^\infty\Psi_\eps(\xi)d\xi<\infty.
\]
\end{lemma}
Then Theorem \ref{thm1} follows easily.
%since $n(t,x,y)$ is globally bounded. 

{\bf Proof of Theorem \ref{thm1} from Lemma \ref{lemma3}.} We first
show that Lemma \ref{lemma3} implies Theorem \ref{thm1}. Observe that
maximum principle implies that
\[
0\le n(t,x,y)\le 1.
\]
Then we have
\begin{eqnarray}
&&  \langle V\rangle _t=\frac{1}{t}\int\limits_D[n_0(t,x,y)-n(t,x,y)]
\frac{dxdy}{H}\nonumber\\
&&\le \frac{1}{t}\int\limits_{-\infty}^{(c_*+\eps)t} dx 
\int\limits_0^H
\frac{dy}{H}n_0(t,x,y)+\frac{1}{t}\int\limits_{(c_*+\eps)t}^{\infty} dx 
\int\limits_0^H
\frac{dy}{H}[1-n(t,x,y)]\nonumber\\
&&\le c_*+\eps +\frac{C_\eps}{t}\label{eq:2.3}
\end{eqnarray}
and Theorem \ref{thm1} follows.$\Box$

\section{Proof of Lemma \ref{lemma1}}

We now prove Lemma \ref{lemma1}. 
%The minimal speed $c_*$ for $\Le=1$ may be defined as follows \cite{Freidlin,FG}. 
Given any vector ${\bf
  z}\in{\Rm}^2$ define the linear operator
\[
L_{\bf z}=\kappa(\nabla-{\bf z})^2-u\cdot(\nabla-{\bf
  z})+\frac{v_0^2}{\kappa}g'(0)
\]
with periodic boundary conditions. The operator $L_{\bf z}$ has a
unique continuous positive eigenfunction $\phi(x,y)$ corresponding to a simple
eigenvalue $\lambda({\bf z})$. The positivity of the eigenfunction is standard, while
setting $\phi=e^\omega,$ it is straightforward to verify that if
$u(x,y)$ is mean-zero and incompressible then
$\lambda({\bf z})\ge\frac{v_0^2}{\kappa}g'(0)$. 
The first expressions for the propagation speed in terms of $\lambda({\bf z})$
were given by Freidlin and Gartner \cite{FG,Freidlin}.
The most convenient for our purpose representation for the minimal speed of propagation
in direction $\ve$ was given by Majda and Souganidis \cite{MS}: 
\[
v(\ve)=\inf_{z>0}\frac{\lambda({z\ve})}{z}.
\]
This expression also gives the asymptotic speed of propagation in
direction $\ve=(e_1,e_2)$ for $\Le=1$ and general initial data
satisfying (\ref{eq:2.2}), (\ref{eq:t0+n0=1}), (\ref{compact}). More
precisely, we have for such data
\begin{equation}\label{eq:limitspeed}
\lim_{t\to\infty} T(t,c\ve t)=
\left\{\begin{matrix}{0,~~c>v(\ve)\cr 1,~~c<v(\ve)\cr}
\end{matrix}\right\}.
\end{equation}
In particular the minimal speed of pulsating traveling front in direction $\ve_1=(1,0)$, that we
denote by $c_*$, is given by
\[
c_*=\inf_{z>0}\frac{\lambda({z\ve_1})}{z}.
\]

Given a number $z>0$ let $\Psi(x,y;z)$ be
the positive eigenfunction of $L_z$ corresponding to
$\lambda(z\ve_1)$:
\[
\kappa\left[\left(\pdr{}{x}-{z}\right)^2+\frac{\partial^2}{\partial
y^2}\right]\Psi-u\cdot\nabla\Psi+u_1z\Psi+\frac{v_0^2}{\kappa}g'(0)\Psi=
\lambda(z)\Psi.
\]
We define a comparison function 
\[
\phi(t,x,y;z)=C_ze^{-z(x-{\lambda(z)t}/{z})}\Psi(x,y;z)
\]
with the constant $C_z>0$ so large that
\[
T_0(x,y)\le \phi(0,x,y;z).
\]
We can find such $C$ since $\Psi(x,y;z)\ge\psi_0>0$, and $T_0$
vanishes for $x\ge L_0$.  The function $\phi$ satisfies a partial
differential equation
\[
\phi_t+u\cdot\nabla\phi=\kappa\Delta\phi+\frac{v_0^2}{\kappa}g'(0)\phi
\]
Recall that $n\le 1$, and $g(T)\le g'(0)T$. Therefore $T$ satisfies
\[
T_t+u\cdot\nabla T\le \kappa\Delta T+\frac{v_0^2}{\kappa}g'(0)T
\]
and so the maximum principle implies that for all $t>0$ we have
\[
T(t,x,y)\le\phi(t,x,y;z)=C_ze^{-z(x-{\lambda(z)t}/{z})}\Psi(x,y;z).
\]
We choose then $z>0$ such that $c_*+\eps>\lambda(z)/z$ and obtain the
conclusion of Lemma \ref{lemma1} since $\Psi(x,y;z)$ is bounded from
above. We remark that the proved bound remains true for any $g(T) \leq
MT$ with some $M>0$ not only KPP type. This observation will extend
our arguments to give Theorem~\ref{thm3}. $\Box$

\section{Proof of Lemma \ref{lemma2}} 

Lemma \ref{lemma2} is proved using
the technique of \cite{CX}. We first prove a local $L^p$-bound on $T$.
\begin{lemma}\label{lemma4}
  For any $\gamma>0$ there
  exists a constant $C>0$ so that for any unit cube $Q\subset \Rm^2$ we have
\[
\int\limits_QT^p(t,x,y)dxdy\le Ce^{(\alpha\gamma+\beta\gamma^2)t}
\]
with the constants $\alpha$ and $\beta$ depending only on $\kappa$ and
$U=\|u\|_{\infty}$.
\end{lemma}
{\bf Proof of Lemma \ref{lemma4}.} 
Let $\phi(t,\vx)$ be a smooth test
function and $F(T,n)$ be smooth in both variables. Then we have
\begin{eqnarray}\label{eq:2.4}
&&\pdr{}{t}\int\limits_D\phi(t,\vx)F(T,n)d\vx=\int\limits_D\phi_tFd\vx\\
&&+ \int\limits_D\phi\left[F_T\kappa\Delta
T+\frac{v_0^2}{\kappa}F_Tg(T)n-F_Tu\cdot\nabla
T+\frac{\kappa}{\Le}F_n\Delta n-\frac{v_0^2}{\kappa}F_ng(T)n
-F_nu\cdot\nabla n\right]d\vx\nonumber \\
&&=\int\limits_D[\phi_t+u\cdot\nabla\phi]Fd\vx+\frac{v_0^2}{\kappa}\int\limits_D(F_T-F_n)g(T)nd\vx+
\kappa\int\limits_D\phi\left[F_T\Delta T+\frac{F_n}{\Le}\Delta
n\right]d\vx\nonumber
\end{eqnarray}
We rewrite the last expression above as follows:
\begin{eqnarray*}
&&\int\limits_D\phi\left[F_T\Delta T+\frac{F_n}{\Le}\Delta
n\right]d\vx=-\int\limits_D\left[F_T\nabla\phi\cdot\nabla T+
\frac{F_n}{\Le}\nabla\phi\cdot\nabla n\right]d\vx\\
&&-\int\limits_D\phi\left[F_{TT}|\nabla T|^2+F_{Tn}\nabla n\cdot\nabla
T+ \frac{F_{nn}}{\Le}|\nabla n|^2+\frac{F_{nT}}{\Le}\nabla
n\cdot\nabla T\right]d\vx
\end{eqnarray*}
and insert it into (\ref{eq:2.4}) to get
\begin{eqnarray}\label{eq:2.7}
&&\pdr{}{t}\int\limits_D\phi
Fd\vx=\int\limits_D\left[\phi_t+u\cdot\nabla\phi+\kappa\Delta\phi\right]Fd\vx+
\kappa\left(1-\frac{1}{\Le}\right)\int\limits_DF_n\nabla\phi\cdot\nabla
nd\vx\\ &&-\kappa\int\limits_D\phi\left[\frac{F_{nn}}{\Le}|\nabla
n|^2+\left(1+\frac{1}{\Le}\right)F_{Tn}\nabla n\cdot\nabla T+
F_{TT}|\nabla
T|^2\right]d\vx-\frac{v_0^2}{\kappa}\int\limits_D[F_n-F_T]g(T)nd\vx.\nonumber
\end{eqnarray}
We choose $F$ in such a way that
\begin{equation}\label{eq:2.5}
F_n\ge 2F_T
\end{equation}
and
\begin{equation}\label{eq:2.6}
\left(1+\frac{1}{\Le}\right)^2F_{nT}^2\le F_{nn}F_{TT}.
\end{equation}
Namely, we can take 
\[
F=(A+n+n^2)e^{\eps T}
\]
with $A$ and $\eps$ satisfying
\[
2\eps(A+2)\le 1,~~~A>5\left(1+\frac{1}{\Le}\right)^2.
\]
Then (\ref{eq:2.5}) and (\ref{eq:2.6}) hold and we obtain from (\ref{eq:2.7}):
\begin{eqnarray}\label{eq:2.8}
&&\pdr{}{t}\int\limits_D\phi
Fd\vx\le\int\limits_D\left[\phi_t+u\cdot\nabla\phi+\kappa\Delta\phi\right]Fd\vx+
\kappa\left(1-\frac{1}{\Le}\right)\int\limits_DF_n\nabla\phi\cdot\nabla
nd\vx\\ &&-\frac{\kappa}{2}\int\limits_D\phi\left[\frac{F_{nn}}{\Le}|\nabla
n|^2+F_{TT}|\nabla
T|^2\right]d\vx-\frac{v_0^2}{2\kappa}\int\limits_DF_ng(T)nd\vx.\nonumber
\end{eqnarray}
We choose the function $\phi$ of the form
\[
\phi(\vx)=\frac{1}{(1+\gamma^2|\vx-\vx_0|^2)^2},
\]
then
\[
|\nabla\phi|\le 2\gamma\phi,~~|\Delta\phi|\le 20\gamma^2\phi.
\]
Let us also denote $U=\|u\|_{\infty}$, then we get from (\ref{eq:2.8})
with an appropriate constant $C>0$:
\begin{eqnarray}
&&\pdr{}{t}\int\limits_D\phi Fd\vx\le
 C\int\limits_D\left[U\gamma\phi+\kappa\gamma^2\phi\right]Fd\vx+
 2\kappa\left(1-\frac{1}{\Le}\right)\int\limits_DF_n\gamma\phi|\nabla
 n|d\vx-\frac{\kappa}{2}\int\limits_D\phi\frac{F_{nn}}{\Le}|\nabla
 n|^2d\vx\nonumber\\ &&\le
 \left(CU\gamma+\kappa\gamma^2\right)\int\limits_D\phi
 Fd\vx+2\kappa \Le \left(1-\frac{1}{\Le}\right)^2\gamma^2
\int\limits_D\phi\frac{F_n^2}{F_{nn}}d\vx. \label{eq:2.9}
\end{eqnarray}
(in the last step we replaced the quadratic expression involving $|\nabla n|$
by its maximum). 
Observe that
\[
\frac{F_n^2}{F_{nn}}=\frac{(1+2n)^2e^{2\eps T}}{2e^{\eps T}}\le 5e^{\eps T}\le
2F
\]
as long as $A\ge 3$. Then we have
\begin{eqnarray}
  \label{eq:2.10}
  &&\pdr{}{t}\int\limits_D\phi Fd\vx\le\sigma\int\limits_D\phi Fd\vx
\end{eqnarray}
with
\[
\sigma=C\left[U\gamma+\kappa\gamma^2\right]+
4\kappa\left(1-\frac{1}{\Le}\right)^2\gamma^2
\]
Therefore we obtain
\[
\int\limits_D\phi Fd\vx\le\Gamma e^{\sigma t}
\]
with
\[
\Gamma=\int\limits_D\phi(\vx)F(T_0(\vx),n_0(\vx))d\vx\le \frac{\pi
  e(A+2)}{\gamma^2}.
\]
Then we have for any unit cube $Q$ and any positive integer $k$:
\begin{eqnarray*}
\Gamma e^{\sigma t}\ge\int\limits_D\phi Fd\vx\ge A\int\limits_D
\phi e^{\eps T}d\vx\ge \frac{A\eps^k}{k!}\int\limits_Q\phi(\vx)T^kd\vx\ge
\frac{A\eps^k}{k!(1+\gamma^2)^2}\int\limits_QT^kd\vx.   
\end{eqnarray*}
Here we have chosen $\vx_0$ in the definition of $\phi$ to be the
center of the cube $Q$. Therefore we obtain
\[
\int_QT^kd\vx\le A^{-1}\Gamma e^{\sigma t}(1+\gamma^2)\eps^{-k}k!
\]
for positive integers $k$. Using interpolation we get for all $p>1$
\[
\int_QT^pd\vx\le A^{-1}\Gamma e^{\sigma t}(1+\gamma^2)\eps^{-p}(p+1)^{p+1}.
\]
This finishes the proof of Lemma \ref{lemma4}.$\Box$

We prove now Lemma \ref{lemma2} using Lemma \ref{lemma4}. Let
$G(t,\vx;\vz)$ be the Green's function of the advection-diffusion equation
\[
\phi_t+u\cdot\nabla\phi=\kappa\Delta\phi
\]
posed in the whole space $\Rm^2$. Then we extend $T_0$ periodically to
the whole space $\Rm^2$ from $D$ and obtain
\begin{eqnarray*}
&&  T(t,\vx)=\int\limits_{\Rm^2}G(t,\vx;\vz)T_0(\vz)d\vz+
\int\limits_0^tds \int\limits_{\Rm^2}G(t-s,\vx;\vz)g(T(s,\vz))n(s,\vz)d\vz\\
&&\le 1+
\int\limits_0^tds \int\limits_{\Rm^2}G(t-s,\vx;\vz)g(T(s,\vz))n(s,\vz)d\vz.
\end{eqnarray*}
The Green's function satisfies a uniform upper bound \cite{Norris}
\begin{equation}\label{norris-bd}
G(t,\vx;\vz)\le \frac{C}{4\pi \nu t}e^{-|\vx-\vz|^2/(4\nu t)}
=\tilde G(t,\vx-\vz)
\end{equation}
with some positive constants $C$ and $\nu$ that depend on $\kappa$ and
$u(x,y)$. Then we get
\begin{eqnarray*}
&&T(t,\vx)\le 1+g'(0)
\int\limits_0^tds \int\limits_{\Rm^2}
\tilde G(t-s,\vx-\vz)T(s,\vz)n(s,\vz)d\vz\\
&&=
1+\frac{Cg'(0)}{4\pi \nu t}\int\limits_0^tds 
\int\limits_{\Rm^2}e^{-|\vx-\vz|^2/(4\nu (t-s))}T(s,\vz)n(s,\vz)d\vz.
\end{eqnarray*}
We split the last integral into sum of integrals over unit cubes $Q_j$
with the cube $Q_0$ centered at the point $\vx$:
\begin{eqnarray*}
&&\int\limits_{\Rm^2}e^{-|\vx-\vz|^2/(4\nu (t-s))}T(s,\vz)n(s,\vz)d\vz
=\sum_j\int\limits_{Q_j}e^{-|\vx-\vz|^2/(4\nu
  (t-s))}T(s,\vz)n(s,\vz)d\vz\\
&&\le \sum_je^{-(\hbox{dist}
    (\vx,Q_j))^2/(8\nu (t-s))}\int\limits_{Q_j}e^{-|\vx-\vz|^2/(8\nu
    (t-s))}T(s,\vz)n(s,\vz)d\vz
\end{eqnarray*}
We use the H\"older inequality with $p>1$ in the integral above:
\begin{eqnarray*}
&& \int\limits_{Q_j}e^{-|\vx-\vz|^2/(8\nu(t-s))}T(s,\vz)n(s,\vz)d\vz 
\le \left[~\int\limits_{\Rm^2}e^{-q|\vx-\vz|^2/(8\nu(t-s))}d\vz\right]^{1/q}
\left[\int\limits_{Q_j}T^p(s,\vz)n^p(s,\vz)d\vz\right]^{1/p}\\
&&\le\left[\frac{8\pi\nu(t-s)}{q}\right]^{1/q}
\left[\int\limits_{Q_j}T^p(s,\vz)d\vz\right]^{1/p}\le
C(\gamma,\nu,q)\Le (t-s)^{1/q}e^{(\alpha \gamma+\beta\gamma^2)t/p}
\end{eqnarray*}
Therefore we have
\begin{eqnarray*}
\int\limits_{\Rm^2}\tilde G(t-s,\vx-\vz)T(s,\vz)n(s,\vz)d\vz\le
\frac{C}{(4\pi\nu(t-s))}(t-s)^{1/q}e^{(\alpha \gamma+\beta\gamma^2)t/p}
\sum_j e^{-(\hbox{dist}(\vx,Q_j)^2/(8\nu (t-s))}
\end{eqnarray*}
Note that if $\vz\in Q_j$ with $j\ne 0$ then
\[
|\vx-\vz|\le 2\hbox{dist}(\vx,Q_j)
\]
and therefore
\begin{eqnarray*}
&&\int\limits_{\Rm^2}\tilde G(t-s,\vx-\vz)T(s,\vz)n(s,\vz)d\vz\le
C(t-s)^{-1/p}e^{(\alpha \gamma+\beta\gamma^2)t/p}\left[1+\int\limits_{\Rm^2}
e^{-|\vx-\vz|^2/(16\nu(t-s))}d\vz\right]\\
&&\le
C(t-s)^{-1/p}e^{(\alpha \gamma+\beta\gamma^2)t/p}\left[1+C(t-s)\right]\le
Ce^{(\alpha \gamma+\beta\gamma^2)t/p}\left[(t-s)^{-1/p}+(t-s)^{1/q}\right].
\end{eqnarray*}
Finally integrating over $s\in[0,t]$ we obtain
\[
T(t,\vx)\le 1+Ce^{(\alpha
  \gamma+\beta\gamma^2)t/p}\left[t^{1/q}+t^{1+1/q}\right]\le
C'e^{(\alpha\gamma+\beta\gamma^2)t/p}.
\]
However, $\gamma>0$ is arbitrary and thus Lemma \ref{lemma2} follows. $\Box$

\section{Proof of Lemma 3.} 
Let us define $W=1-C$. It satisfies a
differential equation
\[
W_t+u\cdot\nabla W=\frac{\kappa}{\Le}\Delta W+\frac{v_0^2}{\kappa}g(T)n.
\]
Lemmas \ref{lemma1} and \ref{lemma2} imply that given any $\eps>0$ we
may find $C_\eps, \lambda_\eps>0$ such that
\[
T(t,x,y)\le C_\eps e^{\eps t},~~
T(t,x,y)\le C_\eps e^{-\lambda_\eps(x-c_\eps t)}
\]
with $c_\eps=c_*+\eps$. Let us define
\[
R(t,x)=\frac{v_0^2}{\kappa}
\min\left(C_\eps e^{\eps t},C_\eps e^{-\lambda_\eps(x-c_\eps t)}\right).
\]
Then we have by the maximum principle
\[
W(t,x,y)\le \Phi(t,x,y)
\]
with the function $\Phi$ satisfying the initial value problem
\begin{eqnarray*}
&&  \Phi_t+u\cdot\nabla \Phi=\frac{\kappa}{\Le}\Delta \Phi+R(t,x)\\
&&  \Phi(0,x,y)=T_0(x,y)\le H(-x+L_0).
\end{eqnarray*}
Here $H(x)$ is the Heaviside function:
\[
H(x)=\left\{\begin{matrix}{ 0, x\le 0\cr 1, x>0\cr}\end{matrix}\right.
\]
and $L_0$ is as in (\ref{compact}).  In the sequel we will assume
without loss of generality that $L_0=0$. We let $\Gamma(t,\vx;\vx')$
be the fundamental solution of
\[
\phi_t+u\cdot\nabla \phi=\frac{\kappa}{\Le}\Delta\phi
\]
and obtain ($\vx'=(x',y')$)
\begin{eqnarray*}
  \Phi(t,\vx)=\int\limits_{\Rm^2}\Gamma(t,\vx;\vx')H(-x')d\vx'+
\int\limits_0^tds\int\limits_{\Rm^2}
\Gamma(t-s,\vx;\vx')R(s,x')=\Phi_1+\Phi_2.
\end{eqnarray*}
We will bound $\Phi_1$ and $\Phi_2$ separately. We have a bound for
$\Gamma(t,\vx;\vx')$ similar to (\ref{norris-bd}):
\begin{equation}\label{norris-2}
\Gamma(t,\vx;\vx')\le \frac{C}{\kappa t}e^{-|\vx-\vx'|^2/(C\kappa t)}
\end{equation}
with the constant $C$ depending on $\kappa$, $u$ and $\Le$. Then we obtain
\begin{eqnarray*}
  \Phi_1(t,\vx)\le \frac{C}{\kappa t}
\int\limits_{\Rm^2}e^{-[(x-x')^2+(y-y')^2]/(C\kappa t)}H(-x')dx'dy'=
C'\int\limits_{-\infty}^{-x/\sqrt{C\kappa t}}e^{-z^2}dz\le 
Ce^{-x^2/(C\kappa t)}
\end{eqnarray*}
for $x>0$. Observe that given any $C>0$ we may choose $\lambda_0>0$
and $B>0$ so that for all $x>0$ and all $t>0$ we have
\begin{equation}\label{eq:lambda-0}
e^{-x^2/(C\kappa t)}\le Be^{-\lambda_0(x-c_\eps t)}.
\end{equation}
Indeed we need to find $\lambda_0$ such that
\[
\min_{x>0,t>0}\left\{\frac{x^2}{C\kappa t}-\lambda_0x+\lambda_0c_\eps
  t\right\}>-\infty.
\]
However, we have for $x>0$, $t>0$
\[
\frac{x^2}{Ct}-\lambda_0x+\lambda_0c_\eps t\ge
\frac{2x\sqrt{\lambda_0c_\eps}}{\sqrt{C}}-\lambda_0x>0
\]
for $\lambda_0$ sufficiently small. Therefore (\ref{eq:lambda-0})
holds for such $\lambda_0$ and thus we have
\begin{equation}\label{eq:bd-phi1}
\Phi_1(t,\vx)\le Ce^{-\lambda_0(x-c_\eps t)}~~\hbox{for $x>0$.}
\end{equation}
Our next goal is to obtain such bound for $\Phi_2$. We use the inequality
(\ref{norris-2}) to get
\begin{eqnarray}\label{eq:2.12}
  \Phi_2(t,\vx)\le C\int\limits_0^t ds\int\limits_{\Rm}
\frac{dx'}{\sqrt{\kappa(t-s)}}e^{-(x-x')^2/(C\kappa(t-s))}R(s,x').
\end{eqnarray}
Recall that $R(t,x)$ is defined by
\[
R(t,x)=\left\{\begin{matrix}{C_\eps e^{\eps t},~~x\le X(t)\cr C_\eps
      e^{-\lambda_\eps(x-c_\eps t)},~~x>X(t)\cr}\end{matrix}.\right.
\]
Here $X(t)$ is defined by 
\begin{equation}\label{eq:x(t)}
\eps t=-\lambda_\eps(X(t)-c_\eps
t),~~X(t)=\left(c_\eps-\frac{\eps}{\lambda_\eps}\right)t.
\end{equation}
We split the integral in (\ref{eq:2.12}) accordingly:
\begin{eqnarray}\label{eq:2.13}
 && \Phi_2(t,\vx)\le C_\eps'\int\limits_0^t ds\int\limits_{-\infty}^{X(s)}
\frac{dx'}{\sqrt{\kappa(t-s)}}e^{-(x-x')^2/(C\kappa(t-s))}e^{\eps s}\\
&&+C_\eps'\int\limits_0^t ds\int\limits_{X(s)}^{\infty}
\frac{dx'}{\sqrt{\kappa(t-s)}}
e^{-(x-x')^2/(C\kappa(t-s))-\lambda_\eps(x'-c_\eps s)}=\Phi_{21}+\Phi_{22}.
\nonumber
\end{eqnarray}
The term $\Phi_{21}$ is bounded as follows:
\begin{eqnarray*}
\Phi_{21}(t,\vx)\le C_\eps\int\limits_0^t ds
\int\limits_{-\infty}^{(X(s)-x)/\sqrt{C\kappa(t-s)}}
{dx'}e^{-x'^2}e^{\eps s}
\end{eqnarray*}
Recall that for $\alpha<0$ we have $\int_{-\infty}^\alpha e^{-x^2}dx\le
e^{-\alpha^2}$. Therefore we have for $x>c_\eps t$:
\begin{eqnarray*}
\Phi_{21}(t,\vx)\le C_\eps\int\limits_0^t ds
e^{\eps s-(X(s)-x)^2/(C\kappa(t-s))}\le
C_\eps\int\limits_0^t dse^{\eps s-(x-c_\eps s)^2/(C\kappa(t-s))}.
\end{eqnarray*}
We use now the inequality (\ref{eq:lambda-0}) to get
\begin{eqnarray}\label{eq:phi21-bd}
 \Phi_{21}(t,\vx)\le C_\eps\int\limits_0^t ds
e^{\eps s-\lambda_0(x-c_\eps s-c_\eps(t-s))}
\le C_\eps e^{-\lambda_0(x-(c_\eps+\eps/\lambda_0)t)}. 
\end{eqnarray}

It remains to bound $\Phi_{22}$. The function $\Phi_{22}$ solves the
initial value problem
\begin{eqnarray*}
  &&\pdr{\Phi_{22}}t=C\kappa\frac{\partial^2\Phi_{22}}{\partial x^2}+
C_2e^{-\lambda_\eps(x-c_\eps t)}H(x-X(t))\\
&&\Phi_{22}(0,x)=0
\end{eqnarray*}
with $H(x)$ being the Heaviside function.  
However, we have using (\ref{eq:x(t)})
\[
e^{-\lambda_\eps(x-c_\eps t)}H(x-X(t))\le e^{\eps t}H(-x+c_\eps t)+
e^{-\lambda_\eps(x-c_\eps t)}H(x-c_\eps t),
\]
and therefore
\[
\Phi_{22}\le \Phi_{23}+\Phi_{24}.
\]
The function $\Phi_{23}$ solves the initial value problem 
\begin{eqnarray*}
&&\pdr{\Phi_{23}}t=C\kappa\frac{\partial^2\Phi_{23}}{\partial x^2}+
C_2e^{\eps t}H(-x+c_\eps t)\\
&&\Phi_{23}(0,x)=0.
\end{eqnarray*}
It is bounded exactly in the same way as $\Phi_{21}$ but with $X(t)$
replaced by $x(t)=c_\eps t$. This gives an upper bound
\begin{equation}\label{eq:phi23-bd}
\Phi_{23}(t,\vx)
\le C_\eps e^{-\lambda_0(x-(c_\eps+\eps/\lambda_0)t)}. 
\end{equation}
The function
$\Phi_{24}$ solves the initial value problem
\begin{eqnarray*}
&&\pdr{\Phi_{24}}t=C\kappa\frac{\partial^2\Phi_{24}}{\partial x^2}+
C_2e^{-\lambda_\eps(x-c_\eps t)}H(x-c_\eps t)\\
&&\Phi_{24}(0,x)=0.
\end{eqnarray*}
It may be bounded from above by
\[
\Phi_{24}(t,x)\le\Psi(x-c_{\eps}t).
\]
Here $\Psi(\xi)$ is a positive solution of
\[
-c_\eps\Psi'=C\kappa\Psi''+C_2e^{-\lambda_\eps\xi}H(\xi).
\]
A general solution of the above ODE is given by
\[
\Psi(\xi)=\Psi_0+\frac{C\kappa}{c_\eps}\Psi_1\left[1-e^{-c_\eps\xi/(C\kappa)}
\right],~~~\hbox{for $\xi<0$}
\]
and for $\xi>0$ 
\[
\Psi(\xi)=\Psi_0+\frac{C\kappa}{c_\eps}\Psi_1\left[1-e^{-c_\eps\xi/(C\kappa)}
\right]+\frac{C_2}{(\lambda_\eps(c_\eps-\lambda_\eps C\kappa)}
\left[e^{-\lambda_\eps\xi}-e^{-c_\eps\xi/C\kappa}\right]
+\frac{C_2}{\lambda_\eps
  C\kappa}\left[1-e^{-c_\eps\xi/C\kappa}\right].
\]
Let us require that
\[
\Psi_0+\frac{C\kappa}{c_\eps}\Psi_1+\frac{C_2}{\lambda_\eps c_\eps}=0.
\]
Then the function $\Psi$ takes the form
\[
\Psi(\xi)=\Psi_0e^{-c_\eps\xi/(C\kappa)}+\frac{C_2}{\lambda_\eps c_\eps}\left(
e^{-c_\eps\xi/(C\kappa)}-1\right)
\]
for $\xi<0$, and
\[
\Psi(\xi)=\left[\Psi_0-\frac{C_2}{\lambda_\eps
    (c_\eps-C\kappa\lambda_\eps)}\right]e^{-c_\eps\xi/(C\kappa)}+
\frac{C_2}{\lambda_\eps (c_\eps-C\kappa\lambda_\eps)}e^{-\lambda_\eps\xi}
\]
for $\xi>0$ (we can always assume that 
the denominator is nonzero by shifting $\lambda_\eps$ 
a little bit). Therefore $\Psi>0$ as long as $\Psi_0$ is non-negative,
in particular we may choose $\Psi_0=0$. Then we have
\begin{equation}\label{eq:psi24-bd}
\Phi_{24}(t,x)\le\Psi(x-c_\eps t),~~~\int\limits_0^\infty\Psi(\xi)d\xi<\infty.
\end{equation}
This, together with (\ref{eq:bd-phi1}), (\ref{eq:phi21-bd}) and
(\ref{eq:phi23-bd}) finishes the proof of Lemma \ref{lemma3}.$\Box$

{\bf Acknowledgement.}  We thank N. Vladimirova for discussions
regarding numerical studies related to this work. This work was
partially supported by ASCI Flash Center at the University of Chicago.
AK was supported by NSF grant DMS-9801530, and LR was supported by NSF
grant DMS-9971742.

\end{document}